\documentclass[epsfile,prc,aps,showpacs]{revtex4}
\usepackage{graphicx}
\usepackage{epsfig}

\begin{document}

\today


\title{Variational calculations for K-few-nucleon systems}
\author{S. Wycech }
\address{Andrzej  So{\l}tan  Institute for Nuclear Studies 00-681 Warsaw, Hoza 69,  Poland
\footnote{wycech@fuw.edu.pl}\\}
\author{A.M. Green }
\address{Helsinki Institute of Physics, P.O. Box 64, FIN-00014, Finland
\footnote{anthony.green@helsinki.fi}\\}

\begin{abstract}
Deeply bound KNN,  KNNN and KNNNN states are discussed. The
effective force exerted by the K meson on the nucleons is
calculated with static nucleons. Next the binding energies are
obtained by solving the Schr\"odinger equation or by variational
calculations.

The dominant attraction comes from the S-wave $\Lambda(1405)$ and
an additional  contribution is due to  $\Sigma(1385)$. The latter
state is formed  at the nuclear peripheries  and absorbs   a
sizable piece  of the binding energy. It also generates new
branches  of quasi-bound states.
 The lowest binding
energies based on  a phenomenological KN input fall into  the
40-80 MeV range for KNN,  90-150 MeV for KNNN and 120-220 MeV for
K$\alpha$ systems. The uncertainties are due to unknown KN
interactions in the distant subthreshold energy region.

\end{abstract}

\pacs{PACS numbers:13.75.Jz,  21.45.+v, 25.10.+s, 25.80.-e,
36.10.Gv } \keywords{Deeply bound kaons}
 \maketitle

\section{Introduction }

 In this paper  a   quantitative understanding of  Kaon-few-nucleon
 quasi-bound states is attempted. In recent years,  the existence of such
  states has been vividly discussed.
 It was initiated by the KEK finding of  peaks in the
  nucleon spectra of K$^-$ absorption in  $^4$He,  \cite{KEK04,KEK05}.
  Additional evidence was  given by the  FINUDA measurement of the
   invariant  mass distribution   of the  $\Lambda p $ produced in
     K$^-$absorption by light nuclei \cite{FIN05}.
The existence of such bound states  have been expected
 as the kaon-nucleon and the kaon-nucleus interactions have been  known
 to be strongly  attractive, \cite{WYC86}. This is now firmly confirmed
 on the basis of kaonic atom data \cite{FRI07}.
 However,  the KEK and FINUDA  experiments indicate unexpectedly   strong
 bindings of the order of  100, 150 MeV in the lightest KNN, KNNN   systems.
These experiments require further confirmation. Also,  the interpretation of
the observed peaks has been  disputed in Refs. \cite{VAL06A}, \cite{VAL06B}
while the initial interpretation is defended in Ref. \cite{AKA07}.

Calculations indicate that such states are expected, albeit these might
be very broad and difficult to detect.
 The first calculations  performed by Akaishi and Yamazaki  in
Ref. \cite{AKA02} were followed by several   subsequent
publications. These calculations  exploited essentially the S -
wave resonant attraction related  to the $\Lambda(1405)$ state.
With an optical model type of approach it was shown that the
K-meson optical potential at the center of small nuclei may be as
strong as 500 MeV generating very strong binding of the meson and
a strong contraction of the few-nucleon systems.  However, to
reproduce the KEK data, these calculations involved some
relaxation of the NN repulsion at short distances which would
allow the existence of strongly bound and very dense systems.
These calculations raise the important question on how to
implement   a realistic   short range NN repulsion in the kaonic
systems.

Another open question is related to  the strength and range  of KN
interactions.  Any mathematical description  of few body systems
requires knowledge of NN and KN off-shell scattering amplitudes.
Those related to NN interactions are controlled fairly well in
terms of modern NN potentials. For  a bound K-meson the amplitudes
needed involve  the subthreshold energy region
\begin{equation}
 \label{1}
 f_{KN}=  f_{KN}( - E_B - E_{\rm{recoil}}),
  \end{equation}
where $E_B$ is the KN  separation energy  and $E_{\rm{recoil}} $
the recoil energy of  the  KN pair relative to the rest of the
system. If the separation energy is as  large as 100 MeV, meson
momenta become $\approx 250 $ MeV/c and  $ E_{\rm{recoil}}$ may be
as large as $40$ MeV. The energies of interest  for $ (- E_B -
E_{\rm{recoil}})$ are then located well below the $\Lambda(1405)$
state. The amplitudes there are strongly attractive  and so when
used in a standard optical potential approach may support very
strong bindings.  One problem that arises  at this stage is of a
technical character. As these amplitudes are energy dependent, it
is hard to account for that in the optical model approach. There
 exists  another, more serious, problem  which  is common to all
approaches.
 As the energies involved are far away from the physical region
tested in  KN scattering, the uncertainties in the KN
scattering amplitudes are sizable. For instance, if  the $\Lambda(1405)$ is a
KN bound state,  then the amplitude far below the resonance is
given not only by the position of the singularity but  to a
greater extent by the Born term, which indicates a strong
dependence on the uncertain interaction range $r_o$.  An old
multi-channel potential model of Ref. \cite{KRZ75} indicates that
the  available scattering data do not allow one to fix the precise
value of $r_o$. This unfortunate situation is still actual. The
$r_o$ is expected to be close to the inverse vector meson mass.
However, even though a change of  20$\%$ in $r_o$ would not affect
the scattering region, it  results in a 30$\%$ change of $f_{KN}$
in the deep subthreshold region. The corresponding uncertainty in
the binding energy  then amounts to $\approx 30$ MeV. As indicated
by few body calculations of Ref. \cite{WEI06}
this problem strongly affects the outcome. The uncertainties  of
$f_{KN}$ require further coherent experimental and theoretical
studies of KN and K-few-N interactions.  It becomes one of the
most important  purposes  of the K  meson physics.

On the other hand, there is  one consequence of Eq.(\ref{1}) which
is model independent. If the binding and recoil are so large the
$f_{KN}$ amplitudes involve the energies below the thresholds of
meson-hyperon decay channels. As a consequence the dominant decay
modes are blocked and the lifetimes of nuclear K meson systems are
determined only by multi-nucleon captures. This  leads to the
expectation that such states may  live long enough to be
detectable, \cite{WYC86}.

There exists  several calculations of KNN  binding energies. These
states  are named K$^{-}$pp although in reality they  correspond
to isospin $I_{NN}=1 $ and total isospin $I_{tot}=1/2$. The first
prediction by Akaishi and Yamazaki lead to $( E_B, \Gamma) =
(48,60) $ MeV \cite{AKA02}. With a similar, molecular type method,
Dote and Weise \cite{DOT07} obtain  $ E_B, < 50 $ MeV and indicate
a strong dependence of this result on the short range NN
repulsion. On the other hand, the recent three body calculations
based on Faddeev or AGS methods  yield larger bindings. Thus
Schevchenko \emph{et al.} \cite{SHE07} obtain $( E_B, \Gamma) =
(55-70, 95-110) $  MeV while Ikeda and Sato \cite{IKE07} calculate
$( E_B, \Gamma) = (\sim 80, \sim 73) $ MeV. Later in the text we
show  that  the discrepancy of these two groups of results is due
to: different  $\Lambda(1405)$ properties, explicit description of
the multiple scattering in decay channels and possibly to an
incompatible treatment of the NN repulsion.

There are two new elements introduced in this paper. First, the  P-wave  interactions
due to $\Sigma(1385)$ have been  indicated as a possible source of the strong binding. Here, these are
introduced explicitly. Second, the stress is put on the  strong  KN  spacial correlations
 induced by the S and P wave resonances.

Leaving aside the interpretation
 of the peaks attributed to bound  KNN  and KNNN systems   the essential  theoretical
 questions are:

$(a)$  What is the   binding mechanism?

$(b)$  Are the technical questions under control?

$(c)$  Can the widths be narrow?

\noindent This paper attempts an  answer to these questions and the  following  results are obtained :

1)  To account properly for the KN force range, short range KN
correlations and the NN repulsion,  a two step calculation is
performed. First a wave function involving strongly correlated K-N
subsystems is found in a fixed nucleon approximation. This step
also allows one  to find potentials due to  the   K meson which
tend to contract the nucleons. Next, these  correlated wave
functions and  contracting potentials are used as the input in
variational calculations for the  K-few nucleon binding. In the
KNN case the binding energy and width are  found by solving the
Schr\"odinger equation.

 2) While the dominant
mechanism of attraction is related to the $\Lambda(1405)$ state,
it is found that another resonant state, the $\Sigma(1385)$,
contributes significantly to the structure of the bound states but
much less to the  binding in KNN and K-few-N systems. In addition
the $\Sigma(1385)$ generates new branches of nuclear states that
could not be generated by the  $\Lambda(1405)$ alone.

3)   The  binding energy is determined to a   large extent by the
attraction and the repulsive core in NN interactions. With the
Argonne NN potential  \cite{ARG95} one obtains the lowest state of
KNN   bound by about 40-80 MeV and a  KNNN state bound by about
90-150 MeV. Moderate dependence on the KN interactions is found,
provided these are constrained by the shape of $\Lambda(1405)$ and
the value of  the  KN   scattering length. However, the position
of $\Lambda(1405)$ itself is not well known and this becomes the
source of a large uncertainty. The effect of $ \Sigma(1385)$ on
the binding energy is limited, In the states bound via
$\Lambda(1405)$ it adds some 5-10 MeV contribution to the KNN
binding and 10-20 MeV to KNNN binding. In this sense  the
suggestions of Ref. \cite{WYC07} are not fully supported. However,
the effect of $ \Sigma(1385)$ on the space structure of deeply
bound  kaonic states  is  strong. The $ \Sigma(1385)$ is formed in
peripheral regions and it absorbs  a large fraction of the total K
meson binding. In consequence the
 radii of these systems are fairly  large and the nucleon densities are comparable to
 those met in the  $^4$He nuclei.

 4)  The problem of uncertainties related to the large
recoil momenta entering Eq.({\ref1}) is only partly removed. Large
kaon momenta are hidden inside the resonant structures. In
principle these may be kept under control with the help of other
experiments. In practice it is not the case.  The other sector of
large momenta, due to the strong binding,  is  partly screened by
the short range NN repulsion. The main consequence is a strong
dependence of the meson  binding energies on the position of the
$\Lambda(1405)$ resonance. In principle the shape of
$\Lambda(1405)$ is tested by the invariant mass distribution in
the decay $\Sigma\pi$ channel. In practice it is not so as  the
relevant energy region  is located close to the $\Sigma\pi$
threshold. In this region the theoretical and experimental
uncertainties are large.

5)  These states are very  broad if the binding energies are less
than 100 MeV. For stronger bindings, which  are possible under the
current values of the KN  parameters the main mesonic decay modes
may be closed. The widths for non-mesonic modes are hard to
calculate and extrapolations from the emulsion data are not very
reliable. New experiments are needed.

A simple physical picture emerges from this approach. The mesons are strongly
correlated to slowly moving nucleons.  The correlations are of the $\Lambda(1405)$
type at large densities,  and of the $ \Sigma(1385)$
type in the peripheries.  Each K,N pair
has a good chance to stay also in the $\Sigma ,\pi$ form. The structure is rather
loose as sizable fractions  of the binding energies are hidden in the  short
ranged correlations.

\section{The KNN bound state }

This section presents an   introduction to the method used in this
work.  Several steps describe the increasing degree of precision
and also the increasing  level of technical complications:

$\bullet$ At first the KNN levels are found within the fixed
nucleon approximation with a simple $S$ wave KN  interaction.

$\bullet$ The nucleon degrees of freedom and NN interactions are
introduced  and  a related Schr\"{o}dinger equation is solved.

$\bullet$ The method is extended to multiple channel situations.

$\bullet$  Both   $S$ and $P$ wave KN interactions are allowed.

Consider  scattering of a light meson on two identical, heavy
nucleons. To begin with, the nucleons are fixed at coordinates
$\textbf{x}_i (i=1,2) $ and the wave function is assumed  to be in
the form

\begin{equation}
\label{a2} \Psi(\textbf{x},\textbf{ x}_1 ,\textbf{x}_2) =
\chi_K(\textbf{x};\textbf{ x}_1, \textbf{x}_2)~
\chi_{NN}(\textbf{x}_1,\textbf{x}_2),
\end{equation}
where  $\textbf{x}$  is the meson coordinate. The notation is
simplified and some possible indices are suppressed. The meson
wave function $\chi_K $ is given by the solution of the multiple
scattering equation
\begin{equation} \label{a3}
 \chi_K(\textbf{x},\textbf{x}_1,\textbf{x}_2)= \chi_K(\textbf{x})^o - \Sigma_i ~ \int \textbf{dy }\frac
{\exp[ip\mid \textbf{x-y}\mid ]}
 {4\pi \mid \textbf{x-y}\mid }~ U_{KN}(\textbf{y},\textbf{x}_i)~
\chi_K(\textbf{y},\textbf{x}_1,\textbf{x}_2)
\end{equation}
obtained with fixed positions of the nucleons. An equation  of
similar structure with a zero range meson-nucleon pseudo-potential
$U$ was  used by Brueckner \cite{BRU} to calculate the scattering
length of  a  meson on  two nucleons. For a  high  energy
scattering it  was extensively discussed by Foldy and Walecka, who
used finite range separable interactions $U$ \cite{FOL69}. With
such interactions  equation  (\ref{a3}) allows for  semi-analytic
solutions in the NN,  and also in  few nucleon cases. Here,  the
method is extended to the bound state problem. One  looks  for
solutions of Eq. (\ref{a3}) with no incident wave $
\chi_K(\textbf{x})^o$. The momentum $p$ becomes a complex
eigenvalue $p(x_i)$ which determines  the energy and width of the
system for given nucleon positions $x_i$.

Equation (\ref{a3}) is written in terms of the Klein-Gordon or
Schr\"{o}dinger  propagator. The difference arises when the
relation of energy and momentum is established. Reasons  of
simplicity, which  will become  clear later,  favor the
non-relativistic relation in the  KN center of mass system. Thus,
the interaction is presented as  $U_{KN} =2\mu_{KN} V_{KN}$ where
$\mu_{KN}$ is the reduced mass. Corrections for relativity may  be
introduced at a later stage. The potential $V_{KN}$  for an $S$
wave interaction
 is chosen in a separable form
\begin{equation}
\label{a4}
 V_{KN}( \bf{x-x_i}, \bf{x'-x_i}) =
\lambda ~\upsilon(\bf{x-x_i})~\upsilon(\bf{x'-x_i}),
\end{equation}
where $\upsilon$  is a form-factor and $\lambda$ is a strength
parameter. The eigenvalue equation is now reduced to
\begin{equation}
\label{a5}
 \chi_K(\textbf{x},\textbf{x}_1,\textbf{x}_2) + \Sigma_i ~ \lambda ~\int \bf{dy}~ \frac {\exp[ip(\textbf{x}_1,\textbf{x}_2)\mid \bf{x-y}\mid ]}
 {4\pi \mid \bf{x-y}\mid } ~\upsilon(\bf{y-x_i})~ \int  \bf{dy'}~ \upsilon(\bf{y'-x_i})~\chi_K( \bf{y'},\textbf{x}_1,\textbf{x}_2) =0.
\end{equation}
 Equation (\ref{a5}) becomes a matrix equation for wave
amplitudes $\psi_i$ defined at each scatterer $i$ by
\begin{equation}
\label{a6}
 \psi_i =  \lambda~ \int \bf{dx}~ \upsilon( \bf{x-x}_i) ~\chi_K( \bf{x},\textbf{x}_1,\textbf{x}_2).
\end{equation}
To find the  equations for  $\psi_i $ one introduces  the off-shell KN
scattering matrices $f$  and  matrix elements of the propagator
\begin{equation}
\label{a7}
 G_{i,j}(\bf{x_i,x_j}) =  \int\bf{dy }\bf{dx } ~ \upsilon( \bf{x-x}_i)~
 \frac {\exp(ik\mid \bf{x-y}\mid )}{4\pi \mid \bf{x-y}\mid
 } ~\upsilon(\bf{y-x}_j).
\end{equation}
The diagonal value,   $G_{i,i}\equiv G$,  determines the meson
nucleon scattering matrix $t$ by the well known (see e.g. Ref.
\cite{FOL69}) relation
\begin{equation}
\label{a7t}
 t(E) = (1+\lambda ~ G )^{-1}~ \lambda
\end{equation}
and this yields the full off-shell scattering amplitude $ f $
\begin{equation}
\label{a7f}
 f(k,E,k') =\upsilon(k)~t(E)~\upsilon(k').
 \end{equation}
Here, $k,k'$ are the initial and final  momenta while the
form-factor $v(k)$ is given by the Fourier transform of
$\upsilon(r)$. The Yamaguchi form $\upsilon(k)=
1/(1+k^2/\kappa^2)$ with a free parameter $\kappa$ will be used in
this paper. At zero momenta and at the threshold  this choice
normalizes $f$ (and $t$) to the scattering length. Unfortunately
for  historical  reasons the standard convention in  the K-N
system is to define the scattering length by
\begin{equation}
\label{a7g}
 a+ib  = - f(k=0,E=0,k'=0)\equiv F(0,0,0)
 \end{equation}
 and the capital $F$ will be used in several places to comply with
 the standard KN parameters.

In order to cast   Eq.~(\ref{a5}) into  a standard multiple
scattering equation  for $\psi_i$ one carries out the following
three steps: 1) Integrates Eq.~(\ref{a5}) over the i-th
form-factor  $\upsilon(\bf{x-x_i})$. 2) Selects the i-th term from
the R.H. side. 3)  Multiplies Eq.~(\ref{a5}) by $ (1+\lambda
G)^{-1}$. In this way the kernel of the  multiple scattering
equation can be expressed in terms of scattering amplitudes $t_i$
at each nucleon $i$ and propagators describing the passage from
the nucleon $i$ to the other nucleon $j$. One now  arrives at a
set of linear equations
\begin{equation}
\label{a8}
 \psi_i +
 \Sigma_{j\neq i} ~ t_j~G_{i,j}~  \psi_j  =0,
\end{equation}
which may be solved numerically. For the
 Yamaguchi form-factors,  propagators  $G_{i,j}$ allow  analytic expressions
\begin{equation}
\label{a9}
 G_{1,2}(r,k) =
\frac{1}{ r}~ \upsilon(k)^2 ~[ \exp(ikr)-\exp(-\kappa r)- r
\frac{\kappa^2 +k^2}{2\kappa} \exp(-\kappa r)] ~\equiv~G (r,k),
  \end{equation}
where $ \textbf{r }= \textbf{x}_2- \textbf{x}_1 $. For the sake of
illustration, the KNN  case is presented  in some detail. The
condition for a bound state with two amplitudes $\psi_i$   leads
to   a pair of  equations
\begin{equation}
\label{a10}
 \psi_1 + t~G ~\psi_2 = 0, ~~~~  \psi_2+t~G~\psi_1 = 0.
  \end{equation}
When the determinant
\begin{equation}
\label{a11a} D=  1-( t~G)^2
 \end{equation}
 is put to zero, the
binding "momenta" $p(r)$  may be obtained numerically. Two
different solutions corresponding to $ 1+ tG = 0 $ or $ 1-tG = 0$
may  exist.  The first solution is symmetric $\psi_2= \psi_1$ and
describes  the meson in the $S$ wave state with respect to the NN
center of mass. The second solution is antisymmetric  $\psi_2= -
\psi_1$ and describes a $P$ wave solution. With the  rank one
separable interaction this  latter solution does not   exist in
the full range of $r$. However, it arises with  the more
complicated rank two interactions discussed later.

Eigenvalues corresponding to unstable quasi-bound states are
obtained in the second quadrant of complex $ p(r)\equiv p = p_R +i p_I$
plane.  In this quadrant  the kernel
\begin{equation} \label{a12}
 tG =f(p) ~[ \exp(-p_Ir) \exp(ip_Rr)-\exp(-\kappa r)~( 1+  r \frac{\kappa^2 - p^2}{2\kappa})]/r
  \end{equation}
is exponentially damped at large distances as required by the
asymptotic form of the bound state wave function $\chi_K$. At
short distances $ G $ is regularized by the KN form-factor.

 If the  scattering amplitude is dominated by a quasi-bound state,
 such as  $\Lambda(1405)$,  the related pole  dominates and
 in some energy region $ f\simeq \gamma^2/(E-E^*)$, where   $\gamma$ is a coupling constant and
  $E^*=E_r-i\Gamma_r/2$ is a  complex $\Lambda(1405)$ binding energy.
  The full  KNN binding   energy,  $V_K$,   is given by  the  equation
  $ 1+ tG= 0 $ which   becomes
\begin{equation}
\label{a13} V_K(r) \simeq  E^* - \gamma^2 G(r,p).
\end{equation}
Since Re $ G(r,p)$ close to the resonance is positive this
solution offers binding stronger than the $\approx$  28 MeV
binding in the $\Lambda(1405)$. Asymptotically, for
$r\rightarrow\infty$ one obtains $ V_K \rightarrow  E^* $, that is
a kaon bound to a nucleon to form a  $\Lambda(1405)$.  This type
of asymptotic behavior occurs in all  of the K-few-nucleon systems
of practical interest.  Hence, the separation energy is understood
here as the separation  of the K-N-N system  into  the
N-$\Lambda(1405)$ system.

The limits $ r \rightarrow0$ in Eqs.~(\ref{a12},\ref{a13}) are
regular. However, a joint limit of zero range KN interactions,
$\kappa \rightarrow \infty$, and $ r\rightarrow0$ is singular and
the KNN system collapses. Therefore, some  care is necessary when
this limit is taken.
 Here,
we stay within a phenomenological approach and the standard
expectation that the range of KN interactions is determined by
vector meson exchange. In equation (\ref{a9}) for $G$ the range of
interactions enters twice, first as a  cutoff at small distances
and second in terms of the form-factor $v(k)^2$. We find  in a
numerical way that these two effects cancel and $fG$ is very
stable within the range $ 3< \kappa < 6 ~fm $.
 As the KN interaction range is very short, but finite,
 the   uncertainties related to the actual value of $\kappa$ are additionally  eliminated  by
the short range repulsion in the  NN systems. This  yields  an
 important stability in  the few-body calculations  described
here.

With  nucleons fixed  at a distance $r$  the eigen-value condition
determines $ p(r)$  which in turn generates the  potential
$V_K(r)$ contracting the NN system to a smaller radius. The form
and strength  of this potential depends on the   form of the
kinetic energy.  In the KN  C.M. system the $  E_{KN}$ energy is
given by
 $ \sqrt{M^2 +p^2 } + \sqrt{m^2 + p^2},$
 where $m,M$ are  masses of the meson and nucleon. The same form is kept  in the large
 nucleon mass $M$ limit.
In the  few nucleon  systems  the
 non-relativistic form of the nucleon energy  is used and the problem of large nucleon
 mass disappears.

The  meson propagator in Eq. (\ref{a3}) is chosen to make the
multiple scattering  equation (\ref{a8}) equivalent to a
differential equation
\begin{equation}
\label{a14}
 [ -  \Delta  +  \sum_i ~ 2\mu_{KN}~ V_{ KN_i }~]~\chi  = p(r)^2
 \end{equation}
and the contracting potential becomes   $V_{K}= p(r)^2/ 2\mu_{KN}$
. The advantage of this choice is discussed in the next section.

\subsection{Schr\"{o}dinger equation }

The solution of  the full KNN bound state problem is given by
equation
\begin{equation}
\label{s15}
 ( - \frac{\Delta_x}{2m}  - \frac{\Delta_1}{2M}- \frac{\Delta_2}{2 M} +  V_{KN1}+V_{KN2}+ V_{NN} ) \Psi = E \Psi.
\end{equation}
The  wave function is assumed in the form
    $\Psi= \chi_K(x,x_i) \chi_{NN}(r)$ as given in  Eq.~(\ref{a2}).
Multiplying Eq. (\ref{s15}) on the left by $ \chi_K$  and
integrating over the  meson coordinate   $x$ one obtains  the
Schr\"{o}dinger equation  for  the NN  wave  function
\begin{equation}
\label{s16}
 \chi_{NN} (r) = \int \textbf{dx } \chi_K( x,x_i) \Psi (x,x_i)
  \end{equation}
in the form
\begin{equation}
\label{s17}
 [ E -  V_K(r) +   \Delta_1 / 2 M +  \Delta_2 / 2 M  - V_{NN}]~\chi_{NN}  + \Delta E_{kin} \chi_{NN} =
 0.
  \end{equation}
    where the last term $\Delta E_{kin}$ is a correction to the  kinetic energies.
This correction is small due to the choice of the meson kinetic
energies. In the Schr\"{o}dinger equation (\ref{s15}) it is given
by the meson mass $m$. On the other hand, to determine the
$\Lambda(1405)$ properties and to solve the  scattering equation
(\ref{a3}) the reduced mass $\mu_{KN}$ is used.  Due to this, the
correction  term $ \Delta E_{kin} $ is of the order of $1/M$. In
addition, the meson wave function   satisfies  the relation
\begin{equation}
\label{s19}
 \Delta_x \chi_K  = \sum_i \Delta_i\chi_K,
  \end{equation}
which may be obtained by partial integration over coordinate $y$ in Eq.~(\ref{a5}).
In this way
\begin{equation}
\label{s18} \Delta E_{kin} \chi_{NN} =-\frac{1}{M} \Sigma_i~  \int
\bf{dx }~\chi_K~\overrightarrow{\partial}_i\chi_K
~\overrightarrow{\partial}_i \chi_{NN},
  \end{equation}
which is  very  small due to angular averaging and sign changes in the derivatives.
In more detail this correction reduces to
\begin{equation}
\label{s18a} \Delta E_{kin} \chi_{NN} =-\frac{2}{M}  \int
d\bf{\xi}~ \frac{\overrightarrow{\xi } \overrightarrow{r}}{\xi
r}~~\frac{G(\bf{\xi}-\bf{r})\partial_{\xi} G(\xi)} { \int
d\bf{\eta}[ G(\bf{\eta}-\bf{r})+G(\bf{\eta})]^2 }  ~  ~\partial_r
\chi_{NN}(r),
  \end{equation}
 and is suppressed by the angular average over $\xi$ and   at large $r$
 by the  small overlap of
 $G(\bf{\xi}-\bf{r})$ and $G(\xi)$.  The $\Delta E_{kin}$  makes a
 contribution  $ \approx 0.2 $ MeV  to  the binding energy.
 Such twice damped, small terms of similar type, arise also
in more involved versions of this calculation. The $\Delta
E_{kin}$ is of the same order but is given by very lengthy
formulas. Since it is very small in comparison to the dominant
uncertainties in $V_K$ it is dropped, leading to a significant
simplification of the variational approach.

As the next step, equation (\ref{s17}) is solved with  an S-wave
interaction based on the more realistic NN potential of Argonne
\cite{ARG95}.  This solution is also compared  to  another,
variational solution with the intention
 of checking the variational method used in heavier systems.
The actual interaction used, in the notation of Ref. \cite{ARG95}, has the
form
\begin{equation}
v(NN)=v^{EM}(NN)+v^{\pi}(NN)+v^{R}(NN),
\label{Wir1}
  \end{equation}
where the electromagnetic part $v^{EM}$ only includes the dominant
term proportional to $F_C(r)$ in Eq.~(4) of Ref. \cite{ARG95},
the OPE term $v^{\pi}$ is given by Eq.~(18) of Ref. \cite{ARG95}
 and the phenomenological short range term $v^{R}$ from Eq.~(20)
with the parameters in Table II - again all from Ref. \cite{ARG95}.
This gives directly the S-wave T=1, S=0 interaction $v(\rm{S-wave, \
T=1, \ S=0})$. However, in the T=0, S=1 deuteron channel, the effect of the tensor
interaction
\newline
 $\upsilon_t(T=0, \ S=1)$ on the central component $
 \upsilon_c(\rm{T=0, \ S=1})$
is incorporated by the closure approximation to give
\begin{equation}
V(\rm{S-wave, \ Deuteron})= \upsilon_c(\rm{T=0, \ S=1})-
\frac{8\upsilon_t(\rm{T=0, \ S=1})^2}{\rm{Den}}, \label{Wir2}
  \end{equation}
where the energy denominator was adjusted to Den=338 MeV to ensure the
correct binding energy of the deuteron.

The precision of variational estimates  for E ( used in the next
sections) may be checked against numerical solutions of the
Schr\"{o}dinger equation. It is about 0.3 MeV, compared with the
overall binding of $\sim 50$ MeV.   The  width  of the state is
calculated as
\begin{equation}
\Gamma/2 =  < \chi_{NN}\mid Im~ V_K \mid \chi_{NN} > .
\label{gamma}
\end{equation}

\subsection{Interactions in the decay channels}

The decay channel $ \Sigma\pi$  coupled to the basic KN channel is
now introduced  explicitly.   The wave function at each scattering
center has two components one in the KN the other in  the  $
\Sigma\pi$  channel. The scattering amplitudes  are  two
dimensional vectors $ \psi_i \rightarrow  [\psi_i^K,~ \psi_i^\pi
$] at each nucleon. Multiple scattering equations given in the
previous section are now changed accordingly. One has
\begin{equation}
\label{t1}
 \psi_1^K + t^{K,K}~G^{K,K}~ \psi_2^K +  t^{K,\pi}~G^{\pi,\pi}~ \psi_2^\pi  = 0
  \end{equation}
\begin{equation}
\label{t2}
 \psi_1^\pi  +G^{\pi,\pi}~t^{\pi,\pi}~ \psi_2^\pi + t^{\pi,K}~G^{K,K}~ \psi_2^K  = 0
  \end{equation}
and   an  analogous  pair with $1\leftrightarrow 2$. The notation
has been changed to describe channel indices  and  the  $2\times 2
$ scattering matrix $\hat{f}$.  The determinant related to these
equations gives   the  complex eigenvalue $p(x_i)$ in the KN
channel. The eigen-equation is now more complicated. Introducing a
new notation in channel indices  $ U^{a,b} =G^{a,a}t^{a,b}$  the
determinant becomes
\begin{equation}
\label{t3} D=
 [(1+ U^{K,K})(1+ U^{\pi,\pi})~
- U^{\pi,K}U^{K,\pi}][(1-U^{K,K})(1-U^{\pi,\pi})~
-U^{K,\pi}U^{\pi,K}].
  \end{equation}
The $D=0$ condition  is more transparent close to the singularity
in  the   case of  a  scattering amplitude given by
\begin{equation}
\label{t4}
 f^{a,b} \approx \frac{\gamma_a \gamma_b} { E - E_o+ i \Gamma/2}.
 \end{equation}
Consistency requires the width to be $ \Gamma/2=
p_\pi(\gamma_{\pi})^2$,    where $p_\pi $ is the momentum in the
decay channel. The singular term (\ref{t4})  permits  one to find
a  solution of Eq. (\ref{t3}) in   a fairly  simple form. It is
presented below in the limit of zero range KN ( and  $ \Sigma\pi$
) force. The binding  energy
\begin{equation}
\label{t5} \emph{Re } E =  E_o  - (\gamma_K)^2\frac{cos(p_Rr)}{r}
\exp(-p_Ir) - (\gamma_{\pi})^2 \frac{cos(p_\pi r)}{r}
\end{equation}
becomes larger than the binding of the resonance  but the  collisions
in the  decay channel indicates oscillations.  This oscillatory
behavior is also seen in the width of the system
\begin{equation}
\label{t6}  \emph{Im}~E = - (\gamma_{\pi})^2 ~ p_\pi~[
 1+\frac{sin(p_{\pi}r)}{p_{\pi}r}] -  (\gamma_K)^2\frac{sin(p_Rr)}{r}\exp(-p_Ir).
\end{equation}
The effect of KN scattering represented by the second  term
enlarges the width as $p_R$ is negative. The contribution from
multiple scattering in the decay channel is sizable in general but
it oscillates and may under some  conditions reduce the total
width. That is an effect of  interference in  the decay channel.
Scattering in the decay channel turns out to be constructive in
the KNN case but it is not necessarily so in some heavier systems.

\subsection{ S and P  wave interactions}

With the  KN  interactions allowed in  both  $S$ and $P$ waves the
scattering equation (\ref{a8})  is a $4\otimes 4$ matrix equation
relating four amplitudes $ \psi_i$. The amplitudes for $S$ waves
are now denoted by $ \psi_1^s ,\psi_2^s$. For $P$ wave
interactions the corresponding amplitudes are vectors. As there is
only one vector in the $NN$ system, the relative separation, the
$P$ amplitudes are  chosen to be  $ \textbf{r}\psi_1^p $ and $
\textbf{r}\psi_2^p $. The scattering is now described by  three
types of propagators $ G^{a,b}$ related to consecutive collisions
in the $(S,S)$ $(S,P)$ and $(P,P)$  waves. The scattering
equations are
\begin{equation}
\label{sp1}
 \psi_1^s ~+ ~f^s ~G^{ss}~\psi_2^s   - ~f^s~G^{sp}~r^2 ~\psi_2^p  =0
\end{equation}

\begin{equation}
\label{sp2}
 \psi_2^s ~+ ~f^s ~G^{ss}~\psi_1^s   + ~f^s~G^{sp}~r^2 ~\psi_1^p  =0
\end{equation}

\begin{equation}
\label{sp3}
 \psi_1^p ~+ ~f^p ~G^{pp}~\psi_2^p   + ~f^p~G^{sp} ~\psi_2^s  =0
 \end{equation}

\begin{equation}
\label{sp4} \psi_2^p ~+ ~f^p ~G^{pp}~\psi_1^p   - ~f^p~G^{sp}
~\psi_1^s  =0,
 \end{equation}
where the propagation in between two $P$ wave interactions is
described by  $   G^{pp} = G^{pp}_O +  r^2 G^{pp}_T$. Indices
numbering  the   nucleons  have been suppressed. The propagator
$G^{ss}$ is given in Eq. (\ref{a9})  and  explicit formulas for $
G^{sp},G^{pp}_O , G^{pp}_T $ may be found  in the appendix. All
these functions are regular in the $ r\rightarrow 0$ limit.  The
determinant $D$ of this system  factorizes into two terms
\begin{equation}
\label{sp5}
 D = D_S~D_P,
\end{equation}
where
\begin{equation}
\label{sp6}
 D_S = ( 1 ~+~ G^{ss}~ f ^s)( 1~-~  G^{pp}~ f ^p )~ - ~ G^{sp}~r^2 ~f^s~ f ^p,
\end{equation}
\begin{equation}
\label{sp7}
 D_P = ( 1~ -~ G^{ss}~ f ^s)( 1~ + ~ G^{pp}~f ^p )~-~G^{sp}~r^2~ f^s~ f^p.
\end{equation}
Let us consider the solution of $D_S=0$ close to the
$\Lambda(1405)$ resonance. It is given by  an  equation analogous
to (\ref{a13})
\begin{equation}
\label{sp8} E =  E^* - b^2~ G^{ss}(r,p(E)~ [ 1 ~-~ \frac{r^2(G^{sp})^2
f^p}{1 - G^{pp} f^p}~].
\end{equation}
The second term in parentheses describes the effect of $P$ wave
interactions. At energies below $\Sigma(1385)$  the amplitude
$f^p$ is negative and  generates an additional attraction.

Isospin  symmetry  simplifies  the algebraic structure of the
scattering equations which are (see next sections) expressed by
appropriate isospin combinations of the isospin KN scattering
amplitudes $f$. Equations (\ref{sp1}-\ref{sp4}) allow for a simple
symmetry of the total KNN wave function. Thus, under the
eigenvalue condition (\ref{sp6}) the coordinate wave function
becomes  symmetric with respect to the exchange of nucleon
coordinates. This condition   allows  solutions in terms of two
amplitudes $\psi^s = \psi^s_1 =\psi^s_2$ and $ \psi^p =\psi^p_1 =
- \psi^p_2 $. Wave functions for the KNN system have the form
\begin{equation}
\label{I6}
 \Psi(\bf{r},\bf{x}) =  \chi_{NN}(r)
 [ G(\bf{x}-\bf{r}/2) + G(\bf{x}+\bf{r}/2)] \psi_s  +
\chi_{NN}(r) \overrightarrow{r} \overrightarrow {\partial}_x [
G(\bf{x}-\bf{r}/2) - G(\bf{x}+\bf{r}/2)] \psi_p,
\end{equation}
where $\bf{x}$ is the meson coordinate in the NN center of mass
system,  $ \chi_{NN}(r)$ is the NN wave function. To make this
formula more transparent the  zero range  force limit is taken.
The two terms  in Eq.(\ref{I6}) follow  the  KN interactions in
$S$ and $P$ waves.   The weight of   the  $P$ wave contribution is
given by
\begin{equation}
\label{I7} \frac{\psi_p }{ \psi_s}  = \frac{ f^p~G^{sp}}{1 -
f^p~G^{pp}},
 \end{equation}
which  becomes dominant close to  the zero of the denominator in this equation.
 At large NN separations it happens almost at the singularity  in $ f_p$.
 In this region  the  lowest energy, symmetric,   solution of $D_S=0 $
is given essentially by the situation  $1-  G^{pp} f ^p  \approx 0
$. Such a solution  exists for  $r\geq 1.6 fm$ in the proper
quadrant of the complex momentum. This implies that, at large
separations,  it is energetically profitable for the KNN system to
exist  in the N $\Sigma(1385)$ configuration, with the nucleon and
$\Sigma(1385)$ weakly repelling each other.  At shorter distances
the condition $ 1 + G^{ss} f ^s \approx 0 $ determines the
attraction generated by $\Lambda(1405)$.  Despite repulsive
effects  of the $P$ wave interaction such a solution yields the
strongest binding, since a large piece of the binding energy is
hidden within  the structure of $\Sigma(1385)$.

 The  KNN system is built on short range KN and NN  correlations
and the  $ \Psi(\bf{r},\bf{x})$ contains  a  large number of
partial waves coupled to  zero total angular momentum.  In Jacobi
coordinates $(r,s)$ the
 $L_r \otimes L_s $  decomposition of  the first term of $ \Psi(\bf{r},\bf{x})$
is mainly  $ S \otimes S $.  Both terms involve  even values of  $L_r $  and
the spin-isospin structure of the  NN  pair is either $ I_{NN}=0 , S_{NN}=1$
or   $ I_{NN}=1 , S_{NN}=0$. Both types of states may be formed.

Other  solutions are determined  by $D_P=0$. This condition allows
amplitudes of different symmetry $\psi^s = \psi^s_1 =- \psi^s_2$
and $ \psi^p =\psi^p_1 =  \psi^p_2 $. In comparison with
Eq.~\ref{I6}, the wave functions for the KNN system now become
\begin{equation}
\label{I8}
 \Psi(\bf{r},\bf{x}) =  \chi_{NN}(r)
 [ G(\bf{x}-\bf{r}/2) - G(\bf{x}+\bf{r}/2)] \psi_s  +
\chi_{NN}(r) \overrightarrow{r} \overrightarrow {\partial}_x [
G(\bf{x}-\bf{r}/2) + G(\bf{x}+\bf{r}/2)] \psi_p
\end{equation}
 are  now antisymmetric in the nucleon coordinates and contain
odd angular momenta  $L_r, L_s $ in the $L_r \otimes L_s$
decomposition. The spin-isospin structure of the  NN  pair is
either $ I_{NN}=0 , S_{NN}=0$ or   $ I_{NN}=1 , S_{NN}=1$.  The NN
interactions in the $ I_{NN}=0 , S_{NN}=0$ states are repulsive
and do not support any KNN  bound states. On the other hand,  $
I_{NN}=1 , S_{NN}=1$ states  may be formed.

The results of  the two previous subsections may be unified. The
notation used in Eq.(\ref{t3}) is now extended  to include the
partial wave index in  channel KN : $U^{p,s}= G^{p,p} t^s ,~
U^{s,p}= G^{s,s} t^p ,~ U^{p,p}= G^{p,p} t^p, U^{s,s}\equiv
U^{K,K}$. The last equivalence indicates that the $P$-wave
multiple scattering is included only in the basic KN  channel. For
the determinant of scattering equations one obtains  $ D= D_S~D_P
$ where now
\begin{equation}
\label{I9} D_S = [(1+ U^{K,K})(1+ U^{\pi,\pi})~ -
U^{\pi,K}U^{K,\pi}][1+U^{p,p}]- [(1-U^{\pi,\pi})U^{p,s}U^{s,p}]
  \end{equation}
\begin{equation}
\label{I10} D_P =
 [(1- U^{K,K})(1- U^{\pi,\pi})~
- U^{\pi,K}U^{K,\pi}][1+U^{p,p}]- [(1+U^{\pi,\pi})U^{p,s}U^{s,p}].
  \end{equation}
The solutions of the  corresponding eigen-value equations retain
the symmetries indicated  in the previous section.

Figure 1 indicates  typical contracting potentials obtained with
an A.Martin  solution (see next section and Ref.~\cite{MAR81}).
Asymptotically these reproduce   the separation  energies of the K
meson hidden either in the $S$-wave $\Lambda(1405)$ or in the
$P$-wave $\Sigma(1385)$.  The real binding is generated   by the
difference $ V_K(r)-  V_K(\infty)$ which jointly with  $V_{NN}$
determines the nucleon  wave function $\chi_{NN}$.

\begin{figure}[ht]
\vspace{0.5 cm}
\includegraphics[width=0.4\textwidth]{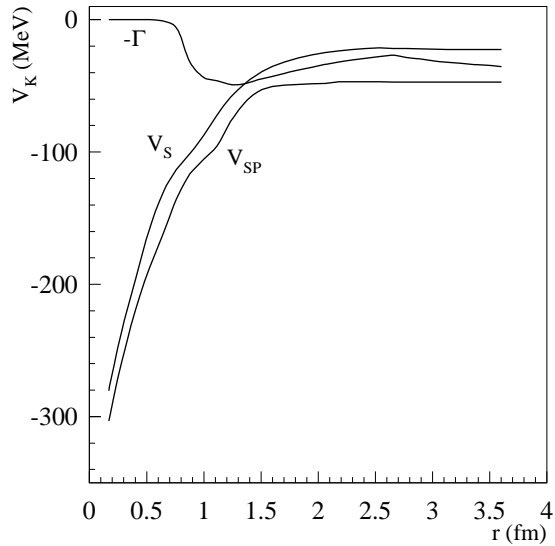}
\caption{Contracting potential $V_K$(r) in the NN, $I_{NN}=1,
I_{KNN}=1/2$
 state, for the  symmetric  meson wave solutions.
 The $V_S$  line  shows  \emph{Re} $V_K(r)$ for
S wave interactions. The $ V_{SP}$  line shows \emph{Re} $V_K(r)$
for S + P wave interactions. The upper curve $-\Gamma$ shows   2
\emph{Im }$V_K(r)$. These results are based on the A. Martin
amplitudes~\cite{MAR81}.}
 \vspace{.0cm} \label{fig:VK}
\end{figure}

\section{KN  interactions}

The coupled multichannel
$KN , \Sigma\pi,\Lambda\pi $ system is  the easiest to describe in terms of
the $\hat{K}$ matrix related to the scattering  matrix
$\hat{T}$ by the algebraic Heitler equation
\begin{equation}
\label{K1} \hat{T} = \hat{K} - \hat{K} i\hat{ Q} \hat{T},
\end{equation}
where $Q$ is a diagonal matrix of channel momenta in the C.M.
system. Early  parameterizations  involved constant  $K$-matrix
elements chosen to fit the scattering data. Later these were
improved by an effective range expansion.  As the data were (and
still are)  poor such
 fits were supplemented by additional consistency conditions
 formulated in terms of dispersion relations \cite{MAR81,MSA69}. Such solutions
 can be  tested above the KN  threshold and to some extent in the $\Sigma\pi $
channel.
For the dominant  isospin $0$ interactions there are two types of solutions.
These are given in  Table \ref{KN} in terms of the inverse
$\hat{M}= \hat{K}^{-1}$ matrix which in turn determines the scattering matrix
\begin{equation}
\label{K2}  \hat{T}^{-1} = \hat{M} + i\hat{ Q}.
\end{equation}

Extrapolations  into  the  complex energy plane display a similar
$\Lambda(1405)$  pole position. However, the physics in both
solutions indicates different interplay of the main KN with the
hyperon pion channels. The position of   the  singularity is given
essentially by  the attractive and, in both cases, large
$K_{KN,KN}$ element. This allows one  to interpret
 $\Lambda (1405)$ as a
$KN$ quasi-bound state.  In principle there exists an alternative possibility - the
 $\Lambda$ as a quark state. If this is the case  it
may  be introduced into the $K$ matrix as  an external  pole in
$\hat{K}\sim 1/(E-E^*)$. However, the scattering  data exclude
such a term or limit it to a very small contribution \cite{MAR81}.
Amplitudes below the KN  threshold may be tested indirectly,
either in the elastic  $\Sigma\pi $  channel or in the $KN
\rightarrow\Sigma\pi $ transitions on bound nucleons \cite{STA87}.
These  reactions support the bound state interpretation but are
not very restrictive on the position of the singularity. In
particular, the analysis of Dalitz and Deloff \cite{DAL91} shows
that several models offer comparable descriptions of the
$\Sigma\pi $   data in the resonance region.  The  $M$ matrix
model given in the DD column of Table \ref{KN} is only slightly
favored by the authors of ref. \cite{DAL91}.
 \footnote{ We thank Andrzej Deloff for supplying
amplitudes of the DD model}. The KWW column in Table \ref{KN}
comes from a quasi-relativistic separable potential model. It
belongs to a second type of solution and was based on the 
B. Martin -- Sakitt solution.
 In this work
we use off-shell extrapolations of both types of solutions.

\begin{table}[h]
\caption{ The semi-phenomenological $I=0$ KN scattering
parameters. First three lines give $M$ matrices at the KN
threshold [fm$^{-1}$]. Next two lines give the  $\Lambda(1405)$
pole position ($E^*$ , $\Gamma/2$) in  the  complex energy plane
[MeV]. The KN scattering length $a_o+ib_o$ and amplitudes at 100
MeV below the threshold \emph{Re}~$F_{-100}+i \emph{Im}~F_{-100}$
are given in units of fm .  The last column KWW$^*$ corresponds to
the
 KWW model modified to change the $\Lambda(1405)$ parameters.
The first solution AM, will be referred to as type one, and the
later solutions BM, DD,  KWW and KWW$^*$ as type two.}
 \vspace{.5
cm} {\begin{tabular}{|l|c|ccc|c|} \hline
 solution & AM\cite{MAR81}  &BM\cite{MSA69} &DD\cite{DAL91}   &KWW\cite{KRZ75}&
KWW$^*$ \\
\colrule
$M_{KN,KN }$&-0.07 &-1.21 & -1.136 & -1.27& -1.27   \\
$M_{KN,\pi\Sigma}$&-1.02 &1.53 &1.254 & 1.50& 1.50 \\
$M_{\pi\Sigma,\pi\Sigma}$& 1.94&-3.05&  -2.205 &-3.05 & -3.05   \\  \hline
$E^*$&1411 & 1415& 1404.9 & 1409 & 1405\\
$\Gamma/2$& 17 & 13 & 26.6 & 22& 24\\ \hline
$a_o$& -1.70 &-1.55 & -1.54 & -1.52 &-1.52\\
$b_o$& 0.68 &0.58 &0.74 &0.60 &0.60\\
$\emph{Re}~F_{-100}$& -2.34  &-1.33  & -8.132 & -2.64 & -5.38\\
$\emph{Im}~F_{-100}$& 0.10 & 0.06 & 1.11 & 0.22 & 0.86\\
\botrule
\end{tabular}
\label{KN}}
\end{table}

\subsection{The   separable off-shell extension}
The three-channel  or two-channel separable model is used here to extend
the phenomenological S-wave KN  interactions
off the energy shell. This method is standard in momentum
space but here  we have already used the
coordinate  representation. In a single  KN channel case the  potential  equivalent to those
of Eq.(\ref{a3}) is  described by
\begin{equation}
\label{S1}  V( k , k' ) =  \lambda
v(k)v(k').
\end{equation}
The Yamaguchi form-factors $v(k) = \kappa^2 /(\kappa^2 + k^2)$ are
convenient to perform explicit analytic calculations in  both
representations. The  related Fourier transforms are of Yukawa
form $ v(r)= \kappa^2\exp(-\kappa r)/ ( 4 \pi r) $ and are
normalized  to delta functions  in the limit of zero range forces.
The off-shell  scattering amplitude becomes
\begin{equation}
\label{S3}
 f_{KN} =\frac{ v(k)v(k')}{ \lambda^{-1} + G(E)}.
 \end{equation}
With the  non-relativistic form of the  kinetic energy  $E_{kin} = q^2/(2\mu_{KN}) $ one obtains
\begin{equation}
\label{S4}
 G(E) =   \int d \bf{\tau} \frac{ v(t)^2}{2\pi^2(\tau^2-q^2)} =
  \frac{ \kappa}{2( 1 - i q/ \kappa )^2} .
 \end{equation}
At the KN  threshold energy  $ E_t=M_K+M_N $, the standard
convention requires  one  to   define the  scattering  length as $
a_o+ib_o = -~f_{KN}(E_t)$. Below  the threshold  $1/\lambda + G$
is forced to have  a zero corresponding to the $\Lambda(1405)$
state. The complex  momentum at this point is
\begin{equation}
\label{S5}
 p_B = [\frac {\lambda_S \kappa^3}{2}]^{1/2} -  \kappa.
  \end{equation}
The next step  in order to improve the absorptive part is to guarantee
that it vanishes below the $\Sigma \pi$ threshold. That is
achieved by scaling  the absorption strength  Im $\lambda $
 by a phase space factor $f_{\varrho} = q_{\Sigma\pi}(E) / q_{\Sigma\pi }( E_t) $
 where $q_{\Sigma\pi}$ is the momentum in  the  decay channel.
 The values  $ \lambda_S = - 0.602\exp(i~0.12~f_{\varrho})~fm $   and $ \kappa = 4.5 fm^{-1}$
 give a good reproduction of the PDG recommended E = 1405 and $\Gamma=50 $ MeV values \cite{DAL91,PDG}.

This one channel amplitude  may serve as a guide, but to describe
finer details and for a  better comparison with the scattering
data one needs  multi-channel separable models. Below, two types
of multi-channel reaction matrices are extrapolated  off the
energy-shell.

$\circ$ One solution has been given by Krzyzanowski $et \ al.$
\cite{KRZ75} in terms of  a  quasi-relativistic multi-channel
separable potential. The $G(E)$ used there  differs from the
solution (\ref{S4}) by an invariant momentum phase space and the
use of quasi- relativistic intermediate meson energies  $E =
q^2/(2M_{N}) + \sqrt{m^2+q^2} $. This solution was motivated by
the early B. Martin and Sakitt \cite{MSA69} K- matrix  (BM in
Table \ref{KN}) and as may be seen in this table it  offers
similar on-shell parameters.

$\circ$ Another solution is based on the commonly used A. Martin's
K-Matrix   \cite{MAR81}. However, in the decay channels this
matrix is not well reproduced by simple rank one  separable
potentials. Instead,  we use the  extrapolation  $ K^{off}(k,E,k')
=  v(k)~K^{on}~ v(k')$,   with the Yamaguchi form-factors  and
$\kappa =$ 4.5 fm$^{-1}$ as obtained above in the one channel
case.  

\subsection{P-wave KN interactions }
To  account for the  P-wave interactions  dominated by  $\Sigma(1385)$
the scattering amplitude of equation (\ref{a7f}) is generalized to
\begin{equation}
\label{P1}
 f_{KN} = f_S  + f_P  = f_S  + 2  {\bf k} {\bf k}'  f_P^{l+}.
\end{equation}
The last term is a consequence  of   the  $j= l+1/2$ total spin of
$\Sigma(1385)$. It involves  an  $l+1$ factor instead of $2l+1$
typical for spin zero situations. The omitted piece contains spin
flip amplitudes and is expected to be small in the few body
context.  The
 $f_P^{l+}$ term is described here  by a separable  single-channel K - matrix
which in the KN channel is given by
\begin{equation}
\label{P2} K( \textbf{k}, \textbf{k'}) =  \gamma_K^2 \frac{
 \textbf{k} \textbf{k'}v_P(k)v_P(k')}{ E- E_o+ i \Gamma_\pi/2},
\end{equation}
where  the form-factor  is
\begin{equation}
\label{P3} v_P(k) =   \frac{\kappa^4 }{(\kappa^2 + k^2)^2},
\end{equation}
and $E_o$ is a phenomenological parameter which determines  the
position of  $\Sigma(1385)$.
The width  $\Gamma_{\pi}$ is strongly energy dependent
\begin{equation}
\label{P4} \Gamma_\pi   = \gamma_{\Lambda\pi}^2 q_{\Lambda\pi}^3 + \gamma_{\Sigma\pi}^2 q_{\Sigma\pi}^3 .
\end{equation}
In these equations  the  $q$'s  are the channel  momenta  and $
\gamma$  are couplings of the resonance to the $\Sigma\pi$ and $
\Lambda\pi$  channels. The latter  are  determined by the
experimental decay width of about 36($\pm 2$) MeV and the
$\Sigma\pi$ branching ratio of  0.13($ \pm$ 0.01 ) \cite{PDG}. For
the coupling to the KN channel the  SU(3) value $ \gamma_{KN}^2/
\gamma_{\Sigma\pi}^2 = 2/3$ is taken.  This value is consistent
with the experimental result of Brown, which yields $ 0.57( \pm
0.18)$ \cite{BRO79}.

The coupling to the  KN channel generates the off-shell scattering
amplitude
\begin{equation}
\label{P5} f_P(\textbf{k}, E, \textbf{k}')  = \frac{2\textbf{k}
\textbf{k}' v_P(k)v_P(k')
 \gamma_{KN}^2 }{ E- E_o+  \gamma_{KN}^2~\int
 \frac{d\bf{\tau} \tau^2 ~v_P(\tau)^2 } {2\pi^2(\tau^2-k_{KN}^2)}+ i~ \Gamma_\pi/2 }
\equiv  \textbf{k} \textbf{k}'v_P(k)f_P(E)v_P(k').
\end{equation}
Let us notice that below the $KN$ threshold the integral in
Eq.(\ref{P5})  deforms significantly the shape of the resonance
profile. The  range parameter $\kappa = 3.8 $  fm$^{-1}$and $ E_o=
1505.2$ MeV are  used  to reproduces the profile tested
experimentally by Cameron {\it et al.}   \cite{CAM78} in the
$\Lambda \pi$  channel. For further applications the coordinate
representation is needed, which is given by equation
\begin{equation}
\label{P6}  f( \textbf{x}, E, \textbf{x}') =  \lambda_P
\overleftarrow{\partial}v_P(x)f_P(E)
v_P(x')\overrightarrow{\partial'}.
\end{equation}
in terms of the Fourier transforms of   form-factors (\ref{P3}).

\section{Few nucleon systems }
The procedure presented  in the KNN section is now extended to
systems consisting of several nucleons.  Practical calculations
are done for three and four nucleons. At first the multiple
scattering equations similar to Eqs. (\ref{a3} - \ref{a5}) in the
previous section are solved in fixed nucleon systems. The  bound
K-meson wave  function  $\chi_K$ is a solution of
\begin{equation}
\label{f1}
 \chi_K(\bf{x},x_1....x_n ) =
- \Sigma_i\Sigma_{\beta}  \int \bf{dy }\frac {\exp(ik\mid \bf{x-y}\mid )}{4\pi \mid \bf{x-y}\mid
 }~ v( \bf{x-x}_i)_{\alpha}~\lambda^{\alpha\beta}~ v(\bf{y-x}_i)_{\beta} ~\chi_K( \bf{y}, x_1....x_n
),
\end{equation}
where indices $ \alpha, \beta $ denote  channels and partial waves
of the  meson-baryon pair. An index related to the  symmetry of
$\chi$ is suppressed. By analogy with Eqs. (6) and (7), equation
(\ref{f1})  may be reduced to a  matrix equation for the wave
amplitudes defined at each scatterer as
\begin{equation}
\label{f2}
 \psi_i^{\alpha} =  \Sigma_{\beta}\int \textbf{dy }~\lambda^{\alpha\beta}~ v(\bf{y-x}_i)_{\beta} ~\chi_K( \bf{y}, x_i ).
\end{equation}
The kernel of  the scattering equation   can be now expressed in
terms of  scattering amplitudes  $f_i^{\alpha, \beta}$  at  each
nucleon $i$ and propagators describing the passage from nucleon
$i$ to another nucleon $j$.  The latter are given by
\begin{equation}
\label{f3}
 G_{i,j}^{\alpha,\beta} =
\int \textbf{dy }\textbf{dx }  v( \textbf{x-x}_i)_{\alpha }
 \frac {\exp(ik\mid \textbf{x-y}\mid )}{4\pi \mid \textbf{x-y}\mid
 } v(\textbf{y-x}_j)_{ \beta}.
\end{equation}
The  procedure explained in Sect. II  leads to a  set of linear equations
\begin{equation}
\label{f4}
 \psi_i^{\alpha} + \Sigma_{\beta\gamma}~
 \Sigma_j ~ f_j^{\alpha,\beta}~ G_{i,j}^{\beta\gamma} ~  \psi_j^{\gamma}  =0 ,
\end{equation}
which are solved numerically. This matrix equation is simplified
as the  $G$'s  are diagonal in channel indices  and  the $f$ are
diagonal in the partial wave index. Still, the corresponding
determinants are complicated algebraic expressions involving
functions $ G$ and $f$.  Numerical  solutions become  a difficult
problem. It has been solved in the following, approximate way. The
determinant consists of many terms which are arranged according to
 the number of collisions.  With up to  four collisions in
the main channel we retain the structure found in the KNN
situation and the determinant $D$ of this system is presented as
\begin{equation}
\label{f5}
 D =  1 +  \Sigma_{pairs}  (D_S~D_P-1)  +  O_{higher~ orders},
\end{equation}
 where  $D_S$ and $D_P$ are defined in Eqs. (\ref{I9}) and (\ref{I10}).  The
main  term is composed of collisions in the  KNN subsystems which
allows  one to keep  track of the wave function symmetry. The
terms of  higher  order in $f$  are  dropped.

The solution of the full K-few-N bound state problem is given by
the equation
\begin{equation}
\label{f6}
 [ E  +      \frac{\Delta_x}{2 m}+    \sum_i \frac{\Delta_i}{2M}- \sum_i
V_{KN_i}-
\sum_{i,j} V_{N_iN_j} ] \Psi(x,x_1,..,x_n)  = 0.
  \end{equation}
 Again we assume the wave function to be
  given by eqs.(\ref{a2}) and (\ref{f1}) i.e.  $\chi_K(x,x_1..x_n) \chi_{N}(x_1,..,x_n))$.
 Projecting   Eq.(\ref{f6}) on $\psi_K$  one obtains  the  Schr\"{o}dinger equation
 for  the few  nucleon wave    function
\begin{equation}
\label{f7}
 \chi_{N} (x_i) = \int \textbf{dx } \chi_K( x,x_i) \Psi (x,x_i)
  \end{equation}
in the form
\begin{equation}
\label{f8}
 [ E -  E^{c}(x_i) +   \sum_i \frac{\Delta_i}{2M}  - \sum V_{NN}]\chi_N  + \Delta E_{kin} \chi_N =0 ,
  \end{equation}
where $\Delta E_{kin}$ is a correction to the nucleon kinetic
energies. As in Eq. (\ref{s18}) it is given by
\begin{equation}
\label{f9} \Delta E_{kin} \chi_N =
  -\frac{1}{M} \Sigma_i  \int \bf{dx }\partial_i\psi_K\partial_i \Psi
  \end{equation}
and as before  turns out  to be very  small due to angular
averaging and sign changes in both the derivatives. As discussed
in the KNN situation this correction
 has been dropped. In deriving equation (\ref{f8}) a
special form (\ref{f1}) of the meson wave function   is used.  As
in Eq. (\ref{s19}) it satisfies the relation
\begin{equation}
\label{f10}
 \Delta_x \chi_K  =\sum_i\Delta_i\chi_K.
  \end{equation}

In the next step, equation (\ref{f8}) is solved  by a variational
method with  the  NN
potential of Argonne \cite{ARG95}. The
trial wave function is of the form
\begin{equation}
\label{f11}
 \chi_N  =\prod_{i,j, i\neq j}[ 1-\exp(-\lambda_c^2
 (\bf{x}_i-\bf{x}_j)^2)]\exp(-\lambda_l
\mid\bf{x}_i-\bf{x}_j \mid) /\mid\bf{x}_i-\bf{x}_j \mid,
   \end{equation}
where $\lambda_c,\lambda_l$ are variational parameters.  This form
is chosen to give the correct asymptotic limit for large $ \mid
\bf{x}_i-\bf{x}_j \mid$ and also gives a vanishing wave function
for small $ \mid \bf{x}_i-\bf{x}_j \mid$ as expected for a strong
repulsion in the NN potential.

\section{Results }
 \subsection{KNN}

In this section the calculations of  KNN levels  are presented.
The sensitivity to  KN input parameters is studied  and the
states of different  symmetry are discussed.

Contracting potentials $V_K(r)$ were calculated with several
solutions for the phenomenological  S-wave $KN$ reaction matrices
presented in Table \ref{KN}. The solutions of  second type may be
well fitted with a rank one separable potential. Here, the
calculations are done with  the  quasi-relativistic model of
\cite{KRZ75}. This model is based on the $K$-matrix of B. Martin
\cite{MSA69}. Numerically  it is fairly close to the separable
potential of Ref. \cite{ALB76}. For the the first type of
solutions in Table \ref{KN} - due to A. Martin \cite{MAR81} no
satisfactory rank one separable approximation is found. This
difficulty  is related to the  large effective range parameters
involved in this $K$ matrix. In order to retain the physics
involved,  a simple off-shell extension is adopted  $ K_{i,j}
\rightarrow  v(k_i)K_{i,j}v(k_j)$. The Yamaguchi form-factors have
been used and the inverse range parameter $\kappa$ was varied over
the range of $3-6$ ~ fm$^{-1}$. The actual value of $\kappa$
affects the multiple scattering via propagator $G_{ss}(k,r)$ of
Eq. (\ref{a9}). Larger values reduce the form-factor $v(k)$ but
enhance the significance of  the small $r$ region in
$G_{ss}(k,r)$. On average these two effects balance very well and
one finds a very weak ($\sim  1$ MeV) dependence of the total
binding energy on the actual value of $\kappa$. The results given
in Fig.1 and in the  tables which follow   are obtained with $
\kappa=4.5$ ~fm$^{-1}$.

 The energies of the most strongly bound KNN, $I_{tot}=1/2$,
$I_{NN}=1$, quasi-bound states are  given in Tables \ref{KNNS1}
and \ref{KNNlow}.  The first table  describes several steps of the
approximation while the second table indicates the dependence of
binding on the KN  input parameters.

The first line  in Table \ref{KNNS1} is determined essentially by
the effects of $\Lambda(1405)$ excitations described  in the
elastic channel only. The second line describes additional effects
due to multiple scattering in the $\Sigma\pi$ channel. The other
two  lines include  the P wave interactions. The energies of
$I_{tot}=1/2$, $I_{NN}=1$ states given by the S wave interactions
and described by multiple scattering in the single,  KN,  channel
span the region of 30-50 MeV. These results are consistent with
the findings of Akaishi and Yamazaki \cite{AKA02} and Dote and
Weise  \cite{DOT07} obtained with different methods. The
differences within this range are due to  a different  KN  and/or
NN input.  As seen in the second and fourth rows, significant
changes arise with the explicit inclusion of the multiple
scattering in the $\Sigma,\pi $ channels. The binding rises by 10
to 20 MeV and the effect of collision broadening is large.

\begin{table}[h]
\caption{ Binding energies and widths  [in  MeV] of the KNN, $I_{tot}=1/2$, $I_{NN}=1$
space-symmetric  states. The results on the left  are based on
 \cite{MAR81}  parameters, the  results on the right  follow KWW \cite{KRZ75}
parameters discussed in Table \ref{KN}.
 The first column  specifies
  the  channels  explicitly involved in the multiple scattering and meson-nucleon partial waves.
$R_{rms}$ is the
 radius mean squared of  the  N-N separation [in fm ]. The last line is obtained with the simplest
 separable potential discussed in the text and the
$I_{KN}=1$ amplitudes  from AM.}
{\begin{tabular}{|c|c|c|c||c|c|c|}
\hline
 solution          & \multicolumn{3}{c|}{AM\cite{MAR81} }  & \multicolumn{3}{c|}{KWW\cite{KRZ75}}\\
 \hline
                  & $E_B$    & $\Gamma$ & $R_{rms}$  & $E_B$& $\Gamma$ & $R_{rms}$   \\ \hline
$KN; ~~S$         &  27    & 36    & 3.1        & 35.5 & 37          & 2.4  \\
$KN,\Sigma\pi;~S $&  37    & 42     & 2.5        & 43.1 & 47          & 2.1  \\
$KN; ~~S,P$       &  49    & 36     & 3.7        & 49.7 & 36          & 3.3  \\
$KN,\Sigma\pi;~S,P$& 52    & 37      & 2.9        & 56.5 & 39         &2.3\\
\hline \hline
 $KN; ~~S$ & 47    & 47      & 2.3        & & &
\\ \hline
\end{tabular} \label{KNNS1}}
\end{table}

 Table  \ref{KNNlow} shows  binding energies obtained with the 
"canonical"  $\Lambda$  pole position  $E = 1405  $ MeV
\cite{DAL91,PDG} which is lower than the position obtained in
other parameterizations. The result given in the second line of
this table  is comparable to the results obtained, with a similar
input,  by Schevchenko {\it et al.}\cite{SHE07}. The latter work
employs a superior Faddeev technique,  but a more detailed
comparison of results is not easy since that  calculation uses  a
rank one separable potential  to describe the   NN interactions.

\begin{table}[h]
\caption{ Binding energies and widths  [in  MeV] of the KNN,
$I_{tot}=1/2$, $I_{NN}=1$ space-symmetric  states. These results
are based on  KWW \cite{KRZ75}  parameters modified to set the
pole of $\Lambda(1405)$  at 1405 MeV and the width at 48 MeV and
given in Table \ref{KN} (KWW$^*$).  The first column specifies
  the channels  explicitly involved in the multiple scattering and the meson-nucleon partial waves.
$R_{rms}$ is the  radius mean squared of  the  N-N separation [in
fm].}
 {\begin{tabular}{|c|c|c|c|}
 \hline
solution  KWW$^*$
  & \multicolumn{3}{c|}{ } \\ \hline
 & $E_B$ & $\Gamma$ & $R_{rms}$ \\ \hline
$KN; ~~S$         &  50    & 51    & 2.05        \\
$KN,\Sigma\pi;~S $&  71    & 85     & 1.81         \\
$KN; ~~S,P$       &  65    & 43     & 2.09        \\
$KN,\Sigma\pi;~S,P$& 78    & 60     & 1.88         \\ \hline
\end{tabular} \label{KNNlow}}
\end{table}

The inclusion of  resonant  P wave interactions increases the
binding by  some 10 MeV. There is some room for different values
as the experimental $K N \Sigma(1385)$ coupling is not certain.
However, the main effect of $\Sigma(1385)$ is a change of
structure in the KNN systems. A sizable portion of the binding
energy is contained in the structure of this resonance. On the
other hand the system is dissolved as the inclusion of P waves
enlarges the NN separation and the formation of $\Sigma(1385)$ is
essentially a peripheral effect. The KN correlations for $ r>1.6 $
fm are mostly of the $\Sigma(1385)$ type. The other effect of P
wave interactions is a formation of additional KNN states. These
are given in  Tables  \ref{KNNS0},\ref{KNNP} and discussed below.

The energies of KNN quasi-bound states with $I_{NN}=0$ given in
Table \ref{KNNS0} are determined essentially by  the
$\Sigma(1385)$ excitations. Let us notice that the result is
unstable against the KN input. The state is still more likely to
exist with a  lower value of the  $\Lambda(1405)$ energy. In any
case it is a very loose structure that might be a quasi-bound or a
virtual state.

\begin{table}[h]
\caption{ Binding energy and width  [in  MeV] of the KNN, $I_{tot}=1/2$, $I_{NN}=0$
space-symmetric  states.
 Results on the left  are based on
 A.Martin parameters  and do not support a bound state (no b.s.). The   results on the right
 follow the KWW parameters.
 This result involves $S+P$ wave interactions and external interactions in
 two meson-nucleon channels. $R_{rms}$ is the
 radius mean squared of  the  N-N separation [in fm]. }
{\begin{tabular}{|c|c|c||c|c|c|}
\hline
 $E_B$    & $\Gamma$ & $R_{rms}$  & $E_B$& $\Gamma$ & $R_{rms}$   \\ \hline
 no~ b.s. & -        &  -         & 47.1 & 36       & $\sim $7  \\  \hline
\end{tabular} \label{KNNS0}}
\end{table}

\begin{table}[h]
\caption{ The binding energy and width [in  MeV]  of the $KNN$
$I_{tot}=3/2$, $I_{NN}=1$ space-asymmetric
 states. In the $NN$ subsystem $^{2S+1}L_J= ^3P_2$.
The last column gives the  radius mean squared of the  N-N
separation, [in fm]. } {\begin{tabular}{|c|c|c|c|c|c|} \hline &
$I_{NNK}$ & $I_{NN}$ & $E_B[MeV]$  & $\Gamma[MeV]$ & $R_{rms}[fm]$
\\ \hline
$K^-nn$ & 3/2 &       1 & 48.5         & 36            & 4.9  \\
\hline
\end{tabular} \label{KNNP}}
\end{table}

The energy of an  asymmetric quasi-bound state  is given in Table
\ref{KNNP}. It is determined essentially by  the   $\Sigma(1385)$
 excitations. The table in Appendix B  indicates that the $K^-nn$ state  has the largest possible
$\Sigma$ component which offers the strongest $\Sigma(1385)-N$
attractive potential. It reaches  a maximum  depth of about $-10$
MeV at $~ 1 ~fm $ distance, but  it is not strong enough to
overcome the NN $P$-wave barrier and generate a quasi-bound state.
To obtain real binding,
 assistance from  the $I_{KN}=1 $  $ S$-wave   state and
the  $NN$ $P$-wave attraction is necessary. Thus, the  $NN$
interactions repulsive at large distances in the $ I_{NN}=0$, $
S_{NN}=0,L_{NN}=1$ waves do not support   bound states. However,
such states may be generated  by $ I_{NN}=1$ , $S_{NN}=1,L_{NN}=1$
interactions. Here, the analysis becomes more subtle as the $NN$
interaction is strongly spin dependent. The energy  given in Table
\ref{KNNP} corresponds to $J=2$ ($^3P_2$) wave in the NN subsystem
where the interaction is the most attractive. This KNN
 system  is large and loosely bound, by about $1.5$ MeV in the
$\Sigma(1385)-N$ configuration. In this calculation the S wave
parameters come from AM \cite{MAR81} and the calculated energy is
uncertain, as the involved $K^-n$ parameters are poorly known. The
experimental detection would be very difficult, nevertheless, a
more precise analysis of the spin and isospin structure of such
states is  of interest in the context of $K^-D$ atoms and will be
performed elsewhere.

\subsection{KNNN  and KNNNN systems}

The discussion of these  systems is limited to the states of the
simplest symmetry. The fixed nucleon model generates  a
contracting potential which in KNNN systems may be, to   a  good
approximation, presented in  the form
\begin{equation}
\label{R1} E^c(R_x, R_y, R_z) = - V_{NNN}[1- C\exp(-\lambda_{s}(
R_x+R_y+ R_z))][\exp(-\lambda_{l}R_x) + \exp(-\lambda_{l}+ R_y)+
\exp(-\lambda_{l}R_z)] - V(\infty) ,
\end{equation}
where $R_x, \ R_y, \ R_z $ are the inter-nucleon distances. The
short range behavior at the triple coincidence may be obtained
analytically and $C= 0.42$, other parameters  being numerical.

With  the KN parameters of Refs.\cite{KRZ75},\cite{MAR81} and
$I_{tot}=0$ the parameters are obtained in the range
$V_{NNN}\sim150-200$ MeV, $\lambda_{s}\simeq4.5$ fm ,
$\lambda_{l}\sim 1.8-1.9$ fm.  For $I_{tot}=1$ one has
$V_{NNN}\sim 50-60$.   $V(\infty)$ is the binding of KN into
$\Lambda(1405)$ in the S wave case or $\Sigma(1385)$ in the S+P
wave case. For $I_{tot}=0$ the corresponding binding energies  are
given on the left side of Table \ref{KNNN0c}. These numbers may be
compared to the simplest version of this model - the $S$ wave
interactions described by the single KN channel - which  produce
91 MeV binding.

The modified version of the KWW model with parameters from Table
\ref{KN} fixed to set the  $\Lambda(1405)$  energy to  1405 MeV
yields
 much stronger contracting forces, $V_{NNN}\sim250-350$ MeV and $\lambda_{l}\sim
2.1-2.3$ fm. The states indicated on the right side in Table
\ref{KNNN0c} are bound very deeply. The basic NNN systems obtained
with our variational wave function are over-bound by about 2 MeV
and this value  has  already been subtracted from the numerical
KNNN energies.

\begin{table}[h]
\caption{ Binding energies and widths  [in  MeV] of the $I_{tot}=
0$, KNNN, space-symmetric  states obtained with the two channel KN
, $\Sigma\pi$ channel multiple scattering formulation.
 The results on the left  are based on the
modified  KWW  \cite{KRZ75} parameters with  $\Lambda(1405)$  energy set to 1405 MeV.
 The first column  specifies meson-nucleon partial waves involved. The widths do not describe
 non-mesic capture modes.  }
 {\begin{tabular}{|c|c|c||c||c|}
\hline
                  & $E_B$    & $\Gamma$  & $E_B$& $\Gamma$    \\ \hline
$S$         &  103    & 29     & 142   & 25  \\
$S+P$       &  119   & 23   & 153     & 21  \\\hline
\end{tabular} \label{KNNN0c}}
\end{table}

\begin{table}[h]
\caption{ Binding energies and widths  [in  MeV] of the $I_{tot}=
1$, KNNN, space-symmetric  states obtained with the two channel KN
, $\Sigma\pi$ channel multiple scattering formulation. These
states are formed as a result of  the $P$ wave interactions with
some assistance of  the $S$ wave attraction. }
 {\begin{tabular}{|c|c|c|}
\hline & $E_B$    & $\Gamma$ \\ \hline
$S+P$       & 63    & 38
\\\hline
\end{tabular} \label{KNNN1c}}
\end{table}

There may exist a number of states in the KNNNN systems. In Table
\ref{KNNNN} one finds only the states with the simplest symmetry,
which involve wave functions symmetric under exchange of the
nucleon coordinates. In the absence of  the K meson the basic
$\alpha$ particle structure is used and only  the $S$ wave NN
interactions are included.  With the tensor interactions described
by Eq.(\ref{Wir2}) this $\alpha$ system is over-bound by about 10
MeV,  and this value has been subtracted from the calculated KNNNN
levels.

\begin{table}[h]
\caption{ Binding energies and widths  [in  MeV] of the KNNNN ,
space-symmetric, $S_{tot}=0$, $I_{tot}=1/2$ states.
 See captions to Table \ref{KNNN0c}.}
 {\begin{tabular}{|c|c|c|c||c|}
\hline
                  & $E_B$    & $\Gamma$  & $E_B$& $\Gamma$    \\ \hline
$S$         &  121    & 25      & 170   & 10\\
$S+P$       &  136    & 20    & 172     & 10 \\\hline
\end{tabular} \label{KNNNN}}
\end{table}

\subsection{Level  widths}

Level widths are calculated as twice the expectation value of $ Im
~V_K$.  The KN  resonant  lifetimes are  strongly energy
dependent, being very  short at the resonance, becoming  longer
below the resonant energies  and staying  infinite below the
thresholds of the decay channels. This trend is reflected by $ Im
~V_K$ in Fig. \ref{fig:VK}.  The  energy dependence contained in
the amplitude $f_{KN}( - E_B- E_{\rm{recoil}})$ of Eq. (\ref{1})
is traded  into the space dependence $f_{KN}(V_K)$. These two
types of averaging give fairly close results provided the final
binding energy is located well above the threshold of the decay
channels. Let us indicate some consequences of this relation.

The states generated by the P wave interactions given in Tables
\ref{KNNS0},\ref{KNNP}, and \ref{KNNN1c} correspond to a fairly
loosely bound $\Sigma(1385)$ and the widths of quasi-bound states
are essentially equal to the width of the $\Sigma(1385)$. This
comes as a result of the peripheral binding  and  weak effects of
the collision broadening in  the $P$ wave resonances. In these
states,  the $V_K$ underestimates slightly the average $ - E_B-
E_{\rm{recoil}}$ and the real widths might be smaller. For the
binding energies in the range 60-90 MeV the  $V_K$ is too small at
large distances and too large at small distances with a reasonably
good average. Let us notice that the level widths generated by the
$S+P$ interactions are smaller than those generated by the $S$
waves alone. This  is due to three factors: the width of
$\Lambda(1405)$ is larger than the width of $\Sigma(1385)$, the
collision broadening in $P$ waves is small and the systems due to
the $S+P$ interactions are less compressed. Let us also notice
very strong sensitivity to the input KN amplitudes. The few
examples of $ Im ~F_{100}$ given in Table \ref{KN}  and the
differences of the widths in the  $ I_{tot}=1/2$, $ I_{NN}= 0$,
KNN state visualize this point.

In cases of very large binding, in the range of 120-200 MeV, one
($\Sigma\pi$) or two ($\Sigma\pi,\Lambda\pi$) decay channels are
blocked and the widths calculated here are over estimated.  Such a
situation is likely to happen in the K-$\alpha$ state. To account
for that effect,  the calculation of the contracting potential was
repeated in an optical potential manner. So, the momenta in the
decay channels $\Sigma\pi,\Lambda\pi$ were related to the binding
energy $E_B$ and  allowed  no outgoing waves. Such a calculation
results in a stronger binding. In the KWW$^*$ model one obtains
binding of 220  instead of the 170  given in Table \ref{KNNNN}.
The real decay width is now given by the multi-nucleon capture
mode.

The multi-nucleon captures are  initiated  by the non-mesic KNN
$\rightarrow$ Y N  mode and the branching ratio for this process
is known from emulsion studies to be about $20\%$ in light nuclei
\cite{ROO79}. The emulsion  data are obtained with  stopped  K
mesons and pertain to the nuclear surfaces. An extrapolation in
terms of a characteristic nuclear densities $\rho$ and two body
phase space $L$
\begin{equation}
\label{R2} \Gamma_{multi} \simeq L ~ \varrho^2 ~ \gamma
\end{equation}
for this  decay was attempted in Ref.\cite{WYC86}. A constant
$\gamma$ may be fixed to the emulsion branching ratio and a 20 MeV
level width in the nuclear matter at 90 MeV binding was obtained.
In the strongly bound, few-body systems the kinematics of the
decay is different since the residual nucleons also take sizable
recoil energies.  Roughly, for a three body decay $ L \sim Q ^2$
where $Q$ is the decay energy. Again, an extrapolation from the
emulsion data in terms of the available phase space and the
involved nuclear density  yields non-mesic capture widths in
 K-$\alpha$ in the  10-30 MeV range.
These  estimates are  somewhat larger than the  12 MeV obtained
for the KNNN system in Ref.\cite{AKA02}.
  Unfortunately   such  extrapolations are   uncertain as the energy dependence in $\gamma$ might be
 large and the $Q$ value is not known. Help  from new experiments
 is necessary to settle these questions.

\section{Conclusions}

In this paper a new method to calculate the deeply bound KNN, KNNN
and KNNNN states has been  presented. The calculation consists of
two steps. First a wave function involving strongly correlated K-N
subsystems is found in a fixed nucleon approximation. This step
also allows one to find potentials due to the   K meson which tend
to contract the  inter-nucleon distances. Next, these correlated
wave functions and contracting potentials are used as  input in
the Schr\"odinger or variational calculations for the
K-few-nucleon binding.

 The lowest binding
energies based on  a phenomenological KN input fall into  the
40-80 MeV range for KNN,  90-150 MeV for KNNN and 120-220 MeV for
K$\alpha$ systems. The uncertainties are due to unknown KN
interactions in the distant subthreshold energy region.

We obtain at least partial answers to the basic  questions
presented in the introduction.

$(a)$  The   binding mechanism:  the dominant mechanism of the
attraction is related to the $\Lambda(1405)$ state. This fact has
been known for a long time. In addition, it is found here, that
the $\Sigma(1385)$ contributes significantly to the structure of
the
 K-few-N bound states but much less to the actual  binding energies.
The bound states are built from the strongly correlated KN
subsystems. At central densities these correlations resemble the
$\Lambda(1405)$ and at peripheries the correlations are made by
the quasi-free $\Sigma(1385)$. Sizable fractions of the binding
energies are contained in the KN correlations. One  consequence is
that even with the strong bindings the  nucleon densities are not
dramatically enhanced  as in Ref.~\cite{AKA02} but can  become a
factor  2-4 larger than the standard nuclear matter density
$\rho_o$.

The presence of $\Sigma(1385)$ resonances  in the  few nucleon
systems generates new states. These are predominantly P-wave
states or states built upon the  P-wave N-N interactions and are
usually broad and loosely bound  developing   long tails built
from the $\Sigma(1385)$ correlation.

$(b)$  The control of technical questions: the choice of
correlated wave functions removes the difficulties related to the
uncertain K-N interaction range and allows one to  use  realistic
N-N interactions.  The  recoil energy of the KN subsystems with
respect to the residual nucleons is described only in  an average
sense. This seems to be the weakest part of this method.

 $(c)$  The widths  are related to the lifetimes of the  $\Lambda(1405)$
 and  $\Sigma(1385)$ enhanced by the collision broadenings.
Under the phenomenological KN interactions ( the $\Lambda $ pole
located at ~$\sim 1412 - 17i $ MeV )  the K-few-N states are
$\sim~40 $ MeV wide.

On the other hand, the models with the $\Lambda $ pole located at
$ \sim 1405 - 25i $ MeV generate   K-few-N states which are more
deeply bound. These may be  either very broad  or quite narrow.
With the binding energies in the 60-80 MeV range ( the KNN case)
very broad  - up to 90 MeV - states are obtained. However,  with
the bindings of 120 MeV (KNNN) or 170 MeV (KNNNN) the single
nucleon decay modes are effectively  blocked. The widths are
strongly reduced and the main decay modes are due to multi-nucleon
K captures. These widths are hard to predict, a simple model
suggested here generate widths of about 20 MeV.

A simple physical picture emerges from this approach. The mesons
are strongly correlated to slowly moving nucleons.  The
correlations are of the $\Lambda(1405)$ type at large densities,
and of the $ \Sigma(1385)$ type in the peripheries.  Each K,N pair
has a good chance to stay also in the $\Sigma ,\pi$ form. The
structure is rather loose as sizable fractions  of the binding
energies are hidden in the  short ranged correlations.

\section*{Acknowledgments}

This work is  supported by the KBN grant 1P0 3B 04229 and  the EU
Contract No. MRTN-CT-2006-035482, \lq\lq FLAVIAnet''. Part of this
work was carried out while one of the authors (SW) was at the
Helsinki Institute of Physics.

\appendix
\section{Propagators}

Several formulas for kernels of multiple scattering equations are
collected  in this appendix.

For  Yamaguchi form-factors, the propagators  $G_{i,j}$  yield
analytic expressions. Thus for two consecutive S-wave interactions
one has
\begin{equation}
\label{22}
 G(r,k)^{ss} =
\frac{1}{4\pi r} v(k)^2 \Big[ \exp(ikr)-\exp(\kappa r)- r
\frac{\kappa^2 +k^2}{2\kappa} \exp(-\kappa r) \Big],
  \end{equation}
where$ \mathbf{r }=  \mathbf{x}_i- \mathbf{x}_j $,
  and the  indices $i,j$ referring to the nucleon sites are
  suppressed.
 For an initial S  wave scattering  followed by a  P wave scattering $G$  becomes a vector
\begin{equation}
\label{22a}
 \textbf{G}(r,k)^{sp} =  \textbf{r}G^{sp}(r,k),
  \end{equation}
\begin{equation}
\label{22b} G^{sp} = \frac{1}{4\pi r} v(k)^2  \Big[ \exp(ikr)
(ikr-1) -\exp(\kappa r)[1 +\kappa r + \frac{r^2(\kappa^2 +
k^2)}{2} + \frac{r^3(\kappa^2 + k^2)^2}{8\kappa}  ]\Big].
\end{equation}
  For two consecutive P wave interactions  the propagator
 is a   tensor  of the form
\begin{equation}
\label{22c}
 \textbf{G}(r,k)^{pp}|_{nm} =  v(k)^2[ \delta_{nm}~G^{pp}_{O}   +   r_{n}
 r_{m}~G^{pp}_{T} ].
\end{equation}
These functions may be expressed in terms of basic integrals
\begin{equation}
\label{22d}
 g_n(r) = \frac{4\pi}{(2\pi)^3} \int  d{\bf p}
\exp(ikr) \{\frac{\kappa^2 }{\kappa^2+p^2}\}^{n}
\frac{1}{p^2- k^2} ,
\end{equation}
 which give by  recurrence
\begin{equation}
\label{23}
 g_{1}(r) = \frac{ \exp(ikr)-\exp(\kappa r)}{ r}, \ \
 g_{2}(r)= g_{1}(r) - \frac{\kappa^2 +k^2}{2\kappa} \exp(-\kappa r)
  \end{equation}
\begin{equation}
\label{23b}
 g_{3}(r)= g_{2}(r) -
\frac{(\kappa^2 +k^2)^2}{8\kappa^3}(1+\kappa r) \exp(-\kappa r), \
\  g_{4}(r)= g_{3}(r) -
\frac{(\kappa^2 +k^2)^3}{16\kappa^5}
(1+ \kappa r + \frac {(\kappa r)^2}{3} )\exp(-\kappa r)
  \end{equation}
and finally
\begin{equation}
\label{24}
 G^{pp}_{O}  = \frac{g_{4}(r)'}{r}, \ \  G^{pp}_T = \frac{ g_{4}(r)''}{r^2} -\frac{ g_{4}(r)'}{r^3}.
 \end{equation}

\section{Isospin symmetry} It is assumed here that the isospin is
conserved in the quasi-bound states of K mesons.  In the lowest
S-wave states of the  KNN systems the isospin wave functions may
be built upon iso-singlet or iso-triplet NN states. From the
experimental point of view, the most interesting one seems to be
\begin{equation}
\label{IT1}
 \Psi^{1/2}_1 =  \{\{NN\}^{1}K\}^{1/2} =
 \sqrt{3}/2\{\{NK\}^{0}N\}^{1/2}+1/2\{\{NK\}^{1}N\}^{1/2},
\end{equation}
where  in $\Psi^{1/2}_1$ the upper index  denotes  isospin $
I_{nucl}$ and the lower index denotes  the spin of the two
nucleons. On the right side the isospin content in the KN
subsystem is given. This state is a mixture of $K^-pp $ and $K^0np$
  and is  frequently  named $K^-pp $
 since it can be experimentally accessed via this entrance
channel. The NN spin in this state is $S=0$ and the effective KN
interaction amplitude obtained from Eq.(\ref{IT1})
becomes
\begin{equation}
\label{IT2}
 f_{KN} =  3/4 f^0_{KN} +1/4 f^1_{KN}.
\end{equation}

Another KNN state of interest is built upon   the NN
iso-singlet
\begin{equation}
\label{IT3}
 \Psi^{1/2}_0 =  \{\{NN\}^{0}K\}^{1/2} =
 -1/2 \{\{NK\}^{0}N\}^{1/2}+\sqrt{3}/2\{\{NK\}^{1}N\}^{1/2} .
\end{equation}
This state is a mixture of $K^-np $ and $K^0nn $  which might be
reached by the $K^-np $   entrance channel. Now the NN spin is
$S=1$ and the effective KN interaction amplitude obtained from
Eq.~\ref{IT3} becomes
\begin{equation}
\label{IT4}
 f_{KN} =  1/4 f^0_{KN} +3/4 f^1_{KN}.
\end{equation}
The S - wave KN interaction in the $\Psi^{1/2}_0 $ state is much
less attractive than in the $\Psi^{1/2}_1 $ state since  the
$\Lambda(1405)$ contribution is reduced. However,  this  is offset
by the strong short range attraction in the NN system due to
 the tensor force. An additional attractive force  is
due to a larger contribution
 from the $\Sigma(1385)$ resonance.

Finally one may have total   isospin 3/2 states of the
type $K^-nn $  or $K^0pp$
\begin{equation}
\label{IT5}
 \Psi^{3/2}_1 =  \{\{NN\}^{1}K\}^{3/2} =   \{\{NK\}^{1}N\}^{3/2}.
\end{equation}
Those states,  involve weakly attractive and  uncertain,
S - wave KN  I = 1 amplitudes. A deeper state can in principle
be built upon the stronger  P - wave interactions. Its
existence and the chances for detection present a  situation
that is  more difficult than the other cases.

For the three nucleon problem we retain the dominant  structure of
the triton and helium isospin symmetry.  The KNNN wave function is
assumed  to be of  the form
\begin{equation}
\label{IT7}
 \Psi^{T} = \frac{1}{\sqrt{2}}\{
 \{\{\{NN\}^{0,1}N\}^{1/2}+\{\{NN\}^{1,0}N\}^{1/2}\}K\}^{T},
\end{equation}
where the pair of  indices denote spin and isospin of the $NN$
pair.

 Re-coupling to the KN system  leads in the total  T=0  state to the relation

\begin{equation}
\label{IT8}
 \Psi^{0} =  \sqrt{1/2}\{\{NN\}^{1} \{NK\}^{1}  +
\{NN\}^{0} \{NK\}^{0}\}^0
\end{equation}
and in this case the  KN  interaction amplitude is
\begin{equation}
\label{IT9}
 f^s =  1/2 f^0_{KN} +1/2 f^1_{KN} .
 \end{equation}
Likewise for the total isospin 1 system
\begin{equation}
\label{IT10}
 f^s =  1/6 f^0_{KN} +5/6 f^1_{KN}
 \end{equation}
These amplitudes are collected into the table.

\begin{table}[h]
\label{IX}
 \caption{Isospin composition of Kaon nucleon scattering amplitudes.
 $I_{tot}$ = total isospin, $I_{nucl}$ = isospin of nucleons,  $f_{i}$ = KN amplitudes of isospin, $i$. }
\begin{center}
\begin{tabular}{|lccr|}
\hline
System  & $I_{tot}$&$I_{nucl}$ & $f_{KN}$   \\
\hline
 KNN & $\frac{3}{2}$  & 1    & $f_{1}$     \\
\hline
 KNN & $\frac{1}{2}$  & 1    & $\frac{3}{4}f_{0}+ \frac{1}{4}f_{1}$     \\
\hline
 KNN & $\frac{1}{2}$  & 0    & $\frac{1}{4}f_{0}+  \frac{3}{4}f_{1}$     \\
\hline
 KNNN & $ 0 $  &  $\frac{1}{2}$    & $\frac{1}{2}f_{0}+  \frac{1}{2}f_{1}$    \\
\hline
KNNN & $1 $  & $\frac{1}{2}$    &   $\frac{1}{6}f_{0}+  \frac{5}{6}f_{1}$ \\
 \hline
 KNNNN & $\frac{1}{2}$& $0$    &   $\frac{1}{3}f_{0}+  \frac{2}{3}f_{1}$ \\
 \hline
\end{tabular}
\end{center}
\end{table}

\section{Three nucleons, S-wave interactions}

The  energy eigenvalue  is obtained by  the  simultaneous solution
of three equations
\begin{equation}
\label{3a}
 \psi^s_1+  G^{ss}_{1,2}f^s \psi_2  + G^{ss}_{1,3}f ^s   \psi^s_3  =0
\end{equation}
\begin{equation}
\label{3b}
 \psi^s_2 +   G^{ss}_{2,3}f ^s \psi^s_3 + G^{ss}_{2,1}f ^s \psi^s_1   =0
\end{equation}
\begin{equation}
\label{3c}
 \psi^s_3+  G^{ss}_{3,1}f ^s\psi^s_1  + G^{ss}_{3,2}f ^s \psi^s_2  =0,
\end{equation}
which require the eigenvalue
condition
\begin{equation}
\label{d3s}
 D_{3s}\equiv 1 -  (f ^s)^2[ G^{ss}_{1,2} G^{ss}_{1,2}+ G^{ss}_{1,3}G^{ss}_{1,3}+G^{ss}_{3,2}G^{ss}_{3,2}] +2(f ^s)^3
 G^{ss}_{1,2}G^{ss}_{2,3}G^{ss}_{3,1} =0.
 \end{equation}
This equation is to be solved numerically. A helpful guide to find  the   symmetry of two physically
meaningful solutions  is  the situation of two equal $NN$ separations  $ r_{12} = r_{13}$.
Dropping  the upper indices one obtains the factorized form
\begin{equation}
\label{3d}
 D_{3s}= (1 -   f~  G_{1,2})( 1+ f~G_{1,2} - 2~f ^2~ G_{1,3}^2).
 \end{equation}
 The first factor corresponds to an antisymmetric solution with the meson sticking
 to two nucleons only. The second factor generates  a solution  symmetric with the interchange
 of nucleons 1 and 2. These solutions are a direct continuation of the two
 solutions obtained in the $KNN$ systems.


\begin{thebibliography}{0}

 \bibitem{KEK04}
 T. Suzuki $ et \ al$., {\it Phys.Lett.} \  {\bf B597}(2004)263 :
 \bibitem{KEK05}
  M. Iwasaki,  {\it Proc.EXA 05, Austr. Acad. Sc. Press, 2005 }p.~93:
\mbox{nucl-ex/07060297}
 \bibitem{FIN05}
 M. Agnello  for FINUDA Collab.
{\it  Phys.Rev.Lett.\/} {\bf 94}(2005)212303
\bibitem{WYC86} S. Wycech, {\it Nucl.Phys.} \  {\bf A 450}, 399c, (1986)
\bibitem{FRI07}
 E. Friedmann and A.Gal, {\it Physics Reports } \ {\bf452} (2007)89
\bibitem{VAL06A}
 E. Oset  and H.Toki {\it  Phys. Rev. \/} {\bf C74}(2006)015207
 \bibitem{VAL06B}
  V.K. Magas, E. Oset, A. Ramos and H. Toki {\it  Phys. Rev. \/} {\bf
  C74}(2006)025206,
\bibitem{AKA07}
 Y. Akaishi and T. Yamazaki,  {\it  Nucl.Phys. \/} {\bf A792}(2007)229

\bibitem{AKA02}
 Y. Akaishi and T. Yamazaki,  {\it  Phys. Rev. \/} {\bf C65}(2002)044005
\bibitem{WEI06}
 W. Weise,  {\it Proc.EXA 05, Austr. Acad. Sc. Press, 2005 }p.~35

\bibitem{KRZ75}
 W. Krzyzanowski, J. Wrzecionko and S. Wycech
  {\it et al}., {\it  Acta.Phys.Pol.\/} {\bf B6}(1975)259
\bibitem{DOT07} A. Dote and W. Weise,  {\it EPJ A }, nucl-th/0701050
{\it Prog.Theor.  Phys. Suppl. \/} {\bf 168}, 593(2007)

\bibitem{SHE07}
 N. Shevchenko, A. Gal  and J. Mares,  {\it  Phys. Rev. \/} {\bf C76}, 044004(2007)

\bibitem{IKE07}
Y. Ikeda and T. Sato,   {\it  Phys. Rev. \/} {\bf C76},
035203(2007)

\bibitem{ARG95}
 R.B. Wiringa   {\it et al}., {\it  Phys. Rev. \/} {\bf C51}(1995)38
\bibitem{WYC07}
   S. Wycech and A.M. Green,
    {\it Proc.EXA 05, Austr. Acad. Sc. Press, 2005 }p.~101,  {\it IJMP}{\bf A22}, 629(2007)


\bibitem{BRU}
 K.A. Brueckner, {\it Phys. Rev.\/} {\bf 89}, 834(1953)

\bibitem{FOL69}
 W.L. Foldy and J.D. Walecka, {\it  Ann.Phys.\/} {\bf 54}(1969)447
\bibitem{MAR81}
 A.D. Martin {\it  Nucl.Phys. } {\bf B94}(1975)413
\bibitem{MSA69}
 B.R. Martin and M. Sakitt,  {\it  Phys.Rev } {\bf 183}(1969)307,
 B.R. Martin in {\it Proceedings of Herceg Novi Summer School 1972. }
B.R. Martin,  {\it  Nucl.Phys. } {\bf B184}(1981)33
\bibitem{ALB76}
 M. Alberg, E. Henley and L. Wilets, {\it Ann. of Phys. } {\bf
 96}(1976)43.


\bibitem{DAL91}
 R.H. Dalitz and A. Deloff, {\it  J.Phys.G } {\bf 17}(1991)289

\bibitem{BRO79}
 O. Brown,  {\it  Nucl.Phys \/} {\bf 129}(1979)1
\bibitem{CAM78}
 W. Cameron {\it et al}. ,  {\it  Nucl.Phys. \/} {\bf 143}(1978)189


\bibitem{PDG}
Particle Data Group,  {\it  Phys.Rev . \/}{\bf D 50}, 1734(1990)

\bibitem{FRI05}
E. Friedmann, A. Gal, J. Mares and  A. Cieply, {\it  Phys.Rev .
\/} {\bf60}, 024314(1990)

\bibitem{FRI06}
 J. Mares, E. Friedmann and  A. Gal, {\it Nucl. Phys. \/} {\bf A 777}, 84(2006)



 \bibitem{STA87}  R. Staronski and S. Wycech, {\it Journ. Phys.} \  {\bf
G13}, (2004)1361 :




\bibitem{AKA07A}
T.Yamazaki and  Y. Akaishi,  {\it  Phys. Rev. \/} {\bf C76},
045201(2007)




\bibitem{ROO79}
R. Roosen {\it et al.}, {\it  Nuovo. Cim. \/} {\bf 49A}, 217(1979)







 \end{thebibliography}
\end{document}